# Some comments on a new type of superconducting gravity wave detector

A. Gulian[a], J. Foreman[b], V. Nikoghosyan[a,c], L. Sica[a], J. Tollaksen[a], S. Nussinov[a,d,*,1]

[a]*Chapman University, Institute for Quantum Studies, Orange CA, 92866 & Burtonsville MD, 20866, USA*
[b]*Independent Researcher, Alexandria VA, 22310, USA*
[c]*Physics Research Institute, National Academy of Sciences, Ashtarak, 0203, Armenia*
[d]*School of Physics&Astronomy, Tel-Aviv University, Ramat-Aviv, 69978, Tel-Aviv, Israel*

*Abstract:* We have recently suggested a new approach and design of an ultra-sensitive gravity wave detector antenna based on superconductivity. The idea was described in a short paper [1]: http://iopscience.iop.org/1742-6596/507/4/042013, in entrees on the arXiv [2]: http://arxiv.org/abs/1111.2655, and at various conferences. Here we would like to explain in a more detailed manner the motivation for and the advantages of our approach.

## I. Introduction and motivation

Only gravity waves (GW) freely traveling across the whole universe can probe its very early, extremely hot and energetic beginning. Indeed it has been suggested that the "B patterned" polarization of the cosmic microwave background radiation recently observed by the BICEP2 collaboration [3], if not due to dust, may be attributed to GW emitted at the time of the primordial inflation. This would offer a rare glimpse of the universe at its earliest, highest energy period.
Gravity waves, predicted by the general theory of relativity, have been indirectly observed in precision measurements of the slowing down of binary pulsars [4]. Direct detection of GW will open completely new windows to the cosmos, and has been the holy grail of experimental and theoretical efforts for several decades. Even the idea of using binary pulsars, not as emitters of GW, but as giant antennas for detecting them via the small transient modulation of their frequencies by passing GW, is being vigorously pursued in the "Binary pulsar timing project" [5]. Recently, it was suggested [6] that the Earth may also serve as an indirect GW detector. In the following we will focus on GW detection in experiments with man-made instruments and in particular on a novel approach to and design of such experiments.

## II. Brief review of some basics: GW sources and existing experiments

There are two distinct interdependent aspects for finding GW: the generation at the astrophysical source and the detection. Historically pulsars were considered to be promising sources. They have very precise known frequencies and directions, allowing a much improved signal/noise ratio. However, the intensity of the periodic GW emitted by pulsars is much weaker than that emitted during transient events, such as mergers of two neutron stars, two black holes, a black hole and neutron star, and also core collapse supernovae. Many of the present efforts therefore target such sources.

A key feature underlying all GW detectors is that the response amplitude in each experiment is linear in the amplitude $h$ of the weak incoming GW. The responses are the strain in Weber's aluminum bar, the phase shift in LIGO, and the current in our suggested detector.

Arguably, an ideal detecting instrument need not absorb energy from the GW. Absent dissipative and thermal effects, the detector is in its ground state. The incoming GW then adiabatically distorts the system: at any given time the system is changed a bit in response to the GW to be the new modified system in the presence of the extra tiny gravitational field. Finally when the GW has passed, both the GW and the system, or the instrument, could return to their original state with no energy transfer.

This is not the case for real detectors which always have some dissipation. It holds for all the above three approaches but is most clear in our case: if we use regular conductors as originally suggested by R. Adler [7] rather than superconductors, then the Ohm resistance quickly dissipates the induced currents. This is also the reason why Weber strove to achieve aluminum cylinders with maximal Q values – *i.e.*, minimal dissipation.

The instruments ideally respond to the amplitude rather than to its square. This is a key to the large distance reach of the detectors. The GW amplitude is given in terms of the Schwarzschild radius of the source, its distance $R$ and its quadrupole deformation $\varepsilon$ (eccentricity) by:

$$h \sim R_{Schw} \varepsilon / R. \qquad (1)$$

This falls only *linearly* with distance $R$. If sensitivity to $h \sim 10^{-24}$ can be achieved we could discover mergers even at distances of many mega-parsecs where many such events should occur. Also supernova in our galaxy happening once in ~ 10-to-30 years should generate GW with similar amplitude. The almost spherical collapse has a small quadrupole moment and only a small fraction of the released energy comes out as GW, which is why the detectability range of collapsing sources is much shorter than that of mergers. The signal of mergers has a "chirping" form with increasing intensity and frequency as the compact stars (or black holes) spiral in. Looking for such a pattern of GW in several widely separated detectors could facilitate discovery. Other coincident signatures of the events leading to GW will verify an eventual discovery. These signatures could be a gamma ray burst in the case of a merger or a neutrino pulse for a core collapse event.

At the present time the main experimental effort is based on laser interferometry. In the LIGO experiment [8] the interfering light beams are reflected $\sim 10^3$ times passing a total distance of ~ 4 thousand km. A more ambitious project LISA, aiming to detect longer wavelength GW and based on similar concepts but with mirrors on far satellites, has been temporarily delayed. Exquisite precision and maximal avoidance of any, even quantum, noise are required. Indeed even for $h = 10^{-23}$ the net extra GW induced phase $\delta\phi = 2\pi\delta L/\lambda$ is only:

$$\delta\phi_{GW} = 2\pi L h / \lambda \approx 6 \cdot 4 \cdot 10^8 (cm) \cdot 10^{-23} / 10^{-4} (cm) \sim 10^{-10} \, radian. \qquad (2)$$

Such a phase shift should be detected thanks to the large number $N \sim 10^{20}$ of coherent photons. Since each $\lambda \sim 1\mu m$ photon has energy of ~ $1eV$ this corresponds to 20 Joules stored in the resonant cavity by the 20 kW laser during the $T \approx 10^{-3}$ seconds GW period. It can barely be observed via a slight brightening of the some side fringes in the interfering unperturbed beams.

### III. Qualitative description of the new idea

All instruments designed to detect the extremely weak GW operate on the classical level where the GW field induces physical strains in solids (Weber's case), directly change the optical length experienced by laser beams (in LIGO), or induce currents in superconductors (our case). The extreme weakness of GW implies that we need large numbers of coherent bosons in the detector. These are phonons in Weber's original approach, photons in LIGO, and the supercurrent of Cooper pairs in our superconductors (SC). It is amusing to note that inside $10^5$ cm$^3$ of our envisioned antenna (consisting of an ordinary, low temperature SC) we have $\sim 10^{27}$ Cooper pairs, more than the photons circulating in the LIGO set-up!

If sub-femtoampere currents can be measured in the device we are suggesting, then GW from the sources targeted by LIGO and even weaker sources can be measured. By addressing various questions which naturally arise we try to show that our approach can be implemented.

The basic set-up consists of two SC materials A and B with different electrochemical and physical properties so that current (*i.e.*, Cooper pair motion) induced by GW can flow only from A to B but not in the reverse B to A direction. This "bimetallic" design is crucial. It plays the role of rectifiers in electrical circuits or ratchets in mechanical systems. These are needed to translate the curl-free gravity field of the incident GW into a circulating current, and generating a current is mandatory if we want to benefit from superconductivity. It is amusing to recall a familiar analog of linearly polarized light. Filtering such light by appropriate crystals can convert it into circular polarization – which does carry net angular momentum along the direction of propagation. This circulating pattern is indeed crucial for yet another reason: if charges moving in response to the GW cannot circulate, they accumulate and the resulting strong electric fields would block further motion and quench the response of our antenna. We will show below that the electromotive force caused by GW in the bimetallic loop in our device is indeed non-vanishing.

We should also emphasize the crucial importance of a quadrupole, as opposed to a dipole field, as only the former can generate physical tidal forces. For the low GW frequencies of interest the amplitude *h* is essentially constant over all the GW antennas considered. Adding a slowly time varying $\delta \vec{g}(t)$ to the Earth's dipole acceleration amounts to a gauge, *i.e.*, frame dependent change of the metric. Such a gravitational field generates the same force on any mass element and by the equivalence principle the same acceleration. Thus, the ground on which the lab is situated, the lab (or the satellite), the table on which our antenna is mounted and the electrons, ions, etc. inside it will all accelerate in the same way and no physical observable results ensue. This is *not* the case for the purely transverse GW corresponding to the +2 and -2 helicity components of the graviton [9]. To see how all this works consider then the basic design of the experiment shown in Fig. 1. For simplicity assume that the GW is incident along the *z*-axis. It is then uniform across the *x-y* plane of the drawing with time varying pattern and intensity. The GW forces acting at consecutive π/2 (3π/2) phases of the period are indicated in Fig. 1 by full (broken) arrows and with dark (white) arrows in materials A(B). The forces acting on A and B are always opposite. During the first half of the period the *x*-components of the GW forces stretch the horizontal bars pushing the right halves to the right and the left halves to the left, and the forces along the *y*-axis contract the two vertical bars by pushing the upper halves down and the lower halves up. The same pattern, but with all forces and resulting motions reversed, occurs in the second half cycle. If A is the same material as B and our antenna consisted of only one superconductor, then all forces would identically cancel and there would be no net current flow. However, the longer arrow associated with metal A as compared with the one associated with metal B indicates that the currents generated in A will be stronger than those generated in B. Consequently there will be a net circular flow in the direction prescribed by the longer A arrows and it will be a periodic alternating current with the period of the GW. Different currents would tend to be induced in A and B due to the different physical properties of the A and B superconductors: different densities of the charge carriers, different masses of ions, and different Young moduli.

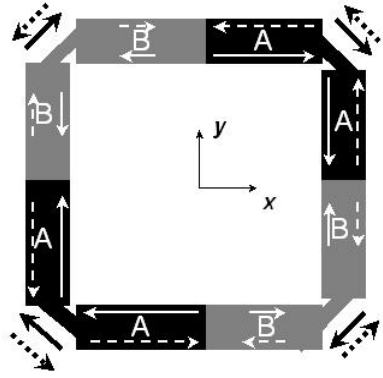

In a simplistic approximation the lattice is treated as completely rigid so that only the electrons and Cooper pairs can move under the influence of the GW. Using the following coordinate deformation: $\Delta x / x = h = h_0 \cos(\omega_0 t) = h(t)$ we find that the electrons /Cooper pairs experience a force proportional to the second time derivative of $mxh(t)$, namely, $mxh_0\omega^2 \cos(\omega t)$. Time integration over half a cycle leads to a velocity $v = \omega x h_0$. This would correspond to an alternating current which at a distance $x$ from the center of the upper BA rod in Fig.1 is:

$$j(x,t) = \omega \sin(\omega t) h_0 e n x = j_0(x) \sin(\omega t); \quad j_0(x) = \omega h_0 e n x \qquad (3)$$

where $e$, $m$, and $n$ are the charge, mass, and number density of electrons/Cooper pairs, $\omega$ and $h_0$ are the frequency and dimensionless amplitude of the GW. The total current would then be obtained by multiplying by the cross-sectional area of our superconducting rods $S$, and using an effective average length $x=L/4$ with $L$ the length of each of the sides of the quadrangle in Fig 1. Since $4L$ is the total length of the four sections, the volume of our superconducting device is $V \sim S4L$ and:

$$I \sim f\omega h e n V / 16 = f\omega h N e / 16 \qquad (4)$$

In an antenna of about hundred liters, we have approximately $N = 10^{27}$. The further "fudge" factor $f < 1$ reflects the fact that the *actual* net circulating current is due to the imbalance between the A and B parts. For an incident GW with $\omega = 2\pi 100$ Hz and $h = 10^{-24}$, we find with $f=1/10$ that the net current has a small but detectable value of *0.4* femtoamperes. However, as we show below, the fact that the action of the GW is transmitted via the much heavier ions greatly enhances the resulting current, so that even $h_0 \sim 10^{-26}$ could be accessible.

## IV. Further, more quantitative analysis

In general, the GW flux emitted by a periodic source is $w = (c^3/16\pi G)\omega^2 h^2 \ ergs \cdot cm^{-2} s^{-1}$ [9] where G is Newton's constant, and the total energy flux falling on a detector of area $L^2$ during one period is $E_f = (c^3/16\pi G)\omega \ h^2 L^2 \ ergs$. A GW incident on four free masses arranged quadrupole-like at the four corners of a square diamond, as indicated in Fig. 2 will adiabatically impart to and retrieve from these masses a kinetic energy $E_{kin} = 2m(L\omega h)^2$. Therefore, the fraction of energy transferred is $\xi_{eff} \sim (32\pi G/c^3)\, m\omega$. This estimate is applicable to any detector showing the critical importance of the total mass responding to the GW. As we will see below, in our device the value of m will be an appreciable fraction of the total mass of the antenna. Still, even for m~100 kg and $\omega$~100s$^{-1}$, $\xi_{eff} \sim 2 \cdot 10^{-30}$. In particular, using the estimate of the GW flux from the Crab Pulsar $w \approx \varepsilon^2$ [10] with eccentricity $\varepsilon$~10$^{-5}$(order of magnitude less than what was suggested by Weinberg ) yields h~10$^{-26}$ and $w \approx 10^{-10} ergs \cdot cm^{-2} s^{-1}$. The geometrical cross section of our antenna is $10^4 cm^2$, and per a half-period, the GW energy available for absorption is $\sim 10^{-8} ergs$, or $10^4 eV$. Given the above efficiency $\xi_{eff} \sim 2 \cdot 10^{-30}$ this yields an energy transfer $\Delta E$ of ~10$^{-26}$ eV, which we are trying to detect. This energy transfer corresponds to a current $I \approx e\sqrt{Sn\Delta E/Lm} \sim 10^{-14} Amp$ for a loop of cross-section S~10$^2$cm$^2$, and L~10$^3$ cm, and the density of charge carriers n~10$^{22}$ cm$^{-3}$. The current for transient sources scaling as h is expected to be larger by a few orders of magnitude, but it lasts for a short period. Prior to estimating the actual current generated in the detector we need to address several issues, the most relevant of which is the possible magnetic blocking of the current.

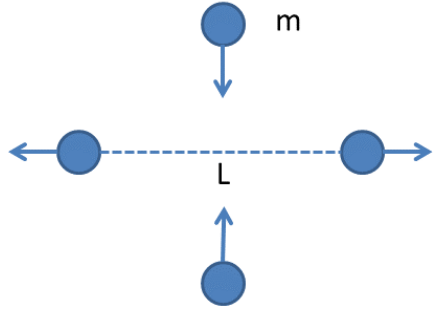

## V. The kinetic and magnetic inertia of our antenna

The kinetic energy of the electrons/Cooper pairs making up the current in our device is

$$K.E. = Nmv^2/2 \qquad (5)$$

where v is the common circulation velocity around the rectangular antenna. Current conservation $div\mathbf{j} + \partial\rho/\partial t = 0$ and the need to avoid any concentration of charges imply that $\mathbf{j}$ is divergence free. For a uniform cross-section and the same carrier density on both A and B, this would indeed require a uniform circulation velocity. While this is not the case in the real system of interest we can use this to obtain a rough approximation of the total kinetic energy of the electrons. Using the parameter values as in Eq. (4) above, we find that in order to achieve the current

$$I=0.4 \times 10^{-15} Amp = nSve = 4000 \ e/sec,$$

with $S=900 cm^2$, $v \sim 5 \cdot 10^{-22} cm/\sec$. Since the mass of $10^{27}$ electrons is one gram, the kinetic energy in the motion of the electrons is:

$$K.E. \sim 10^{-43} ergs. \qquad (6)$$

This is an incredibly small energy, comprising $\sim 10^{-19}$ of the energy $\hbar\omega$ of a single graviton with frequency $\omega \sim 10^3 \ radian/\sec$. What fraction of the GW energy incident on our antenna needs to be adiabatically transferred in order to achieve the above $10^{-43} ergs$? If $10^{51}$ ergs is emitted in GW in a merger event at a distance of 100 mega-parsec$= 3 \cdot 10^{26} cm$, then the integrated fluence on the $\sim 600 \ cm^2$ area of our antenna is $\sim 0.6 \ ergs$. Only $10^{-44}$ of this needs to be transformed into kinetic energy of the current.

However the current induces a magnetic energy in the circuit

$$W_{Magnetic} = \int \frac{d^3r}{8\pi} B^2 \ (cgs) \sim (1/2) L I^2 \ (mks) \ , \qquad (7)$$

where $L$ is the inductance. With the parameters used in Eq (4)

$$W_{Magnetic} \sim 10^{-31} ergs, \qquad (8)$$

which is $\sim 10^{13}$ times larger than the kinetic energy of the electrons. Since this kinetic energy is $\sim mv^2$ and also the magnetic energy is $\sim v^2$ this implies the effective inertia of the electrons is $10^{13}$ larger. This then implies a corresponding decrease of the induced current.

To reduce this large magnetic energy and avoid "magnetic blocking" we use the "spaghetti design" illustrated in Fig 3. Instead of the blocks of A and B material in the basic design of Fig. 1 we use many A and B type superconducting wires. Thus a cross-section perpendicular to the x-direction of the upper part of Fig. 1 would reveal a checker-board pattern of densely packed alternating A and B type thin superconducting filaments. The black squares correspond to A type superconductors each of which is boarded by four neighboring B type superconducting wires indicated by light squares. Each of the A wires is then fused further down steam along the x-direction with a B wire which in turn fuses into an A type and then into a B wire. Thus each black square of Fig. 3 represents an extremely thin version of the complete circuit shown in the original design in Fig 1. The incoming GW will induce in each of these circuits a clock-wise circulating current as depicted in Fig. 1.

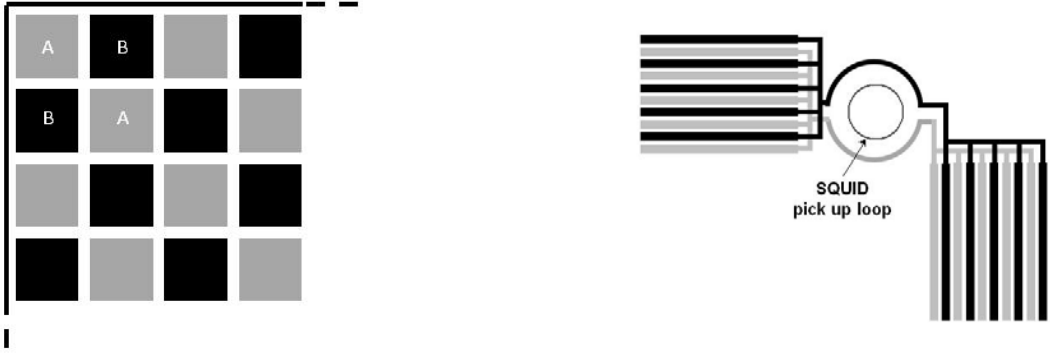

FIG. 3. Left: Cross sectional view of the spaghetti structure. Right: Schematics of the spaghetti interconnection on a planar layer with non-invasive current measurement via a SQUID. Since the currents in materials *A* and *B* flow in opposite directions, the two semi-circles constitute a full current loop for the SQUID pick up.

Likewise each of the white squares in Fig. 3 represents an inverted A → B and B→A circuit in which the same GW will induce a current of the same magnitude but circulating in the opposite sense. For $N_{sp}$ such thin circuits each carrying a current $i = I/N_{sp}$ and each having an area $a \sim S/N_{sp}$ with $I$ and $A$ the total area and total current and all of the same length $\approx L$, the new total magnetic energy becomes:

$$W^{new}_{Magnetic} \sim N_{sp} \cdot a \cdot L \sim W_{Magnetic}/N_{sp} \tag{9}$$

Using $N_{sp} = 10^{10}$ such wires decreases the magnetic energy by this factor. Still $W^{new}_{Magnetic}$ is $10^3$ times larger than the kinetic energy and will reduce the current by a factor of $10^3$. We show in the next section that we greatly underestimated the current and we can afford such a reduction. It is important to note that currents do not penetrate into the bulk of the superconductor by more than $\lambda_{London}$ - the London penetration length which is ~ few hundred angstroms. Thus the spaghetti design where each wire has a radius of $L/(2N_{sp}^{1/2}) \sim 0.1$ micron for a *30 cm* wide antenna with $10^{10}$ wires, makes much better use of the total cross-sectional area of the antenna[2].

Finally in order to read out the net total currents we need to combine, as illustrated in the right portion of Fig. 3, all the currents from the A wires and separately all the currents from the B wires.

### VI. Electrons in SC under GW: *Quo Vadis*?

In this section we discuss the currents induced by GW when lattice deformations are accounted for. A related issue arose long ago and was extensively investigated. Consider a conducting cylinder of height *H* standing on a horizontal table. The gravitational force *mg* on the electrons in the metal is expected to drag them down. The extra negative charge at the lower parts then generates an *E*-field $E = -mg/e$ in a new equilibrium state adjusted to Earth's gravity. The electric potential difference between the top and bottom of the cylinder is thus expected to be

$$V = E \cdot H = -gHm/e. \tag{10}$$

---

[2] The small crosses sectional area of the spaghetti also helps ameliorate the apparent need to conserve the threading magnetic flux which otherwise for the case of a thick superconductor, would have been quantized. Furthermore, the reversed currents in neighboring spaghettis dramatically diminish the magnetic flux.

The negative sign corresponds to the negative charge of the electron. The more careful discussion below indicates that the induced *E* and *V* are larger by a factor $M/m \sim (1-4) \cdot 10^5$. *M* is the mass of the ions in SC materials with typical atomic number ~ 50-200. Also the sign of the electric field is typically reversed. Roughly speaking the metal is squeezed more at the bottom of the bar and the higher density of electrons there generates an electro-chemical potential of the opposite sign of Eq. (10).

To qualitatively understand this we use a simple model, as indicated in Fig 4.

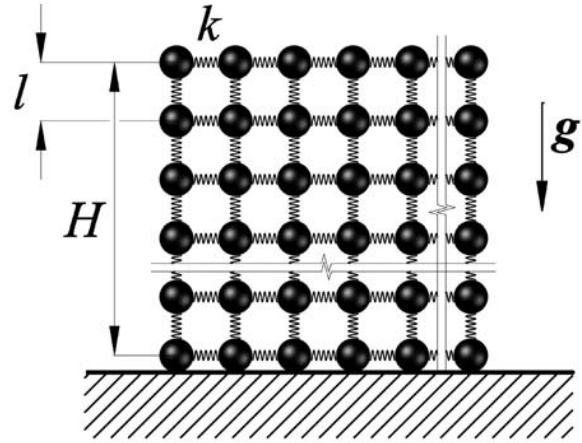

Fig. 4. A schematic description of a lattice. The black circles indicate the ions, and the wavy lines represent the approximately harmonic forces between them.

Each inter-ion bond is approximated by a harmonic spring of equilibrium length *l*= lattice unit ~ 0.1-0.2 nm and spring constant *k*. The weight *(H/l)Mg* of the *H/l* ions over a vertical spring at the bottom of the cylinder compresses the spring by:

$$\delta L/L = F_{gravity}/(kl) = HMg/(l^2 k). \tag{11}$$

We implicitly assumed that the table on which our cylinder stands is infinitely rigid. The springs at the top of the cylinder are not compressed and the inter-ion separation is unchanged there. The increased electron density at the bottom then changes the Fermi energy

$$E_{Fermi} = E_F \sim \hbar^2 n^{2/3}/m, \tag{12}$$

where $n \propto 1/l^3$ is the number density of the electrons, by:

$$\delta E_F / E_F \sim (2/3)\delta n/n \sim -2\delta l/l = -2HMg/(l^2 k). \tag{13}$$

The harmonic approximation $kl^2/2$ for the energy $\sim e^2/l$ required to break the inter-ion bond by extending it from $l$ to $2l$ implies $k \sim e^2/l^3$. Combined with the previous two equations we have $\delta E_F \sim 2MgH\hbar^2/(me^2l)$. Using $l \sim (2-4)a_{Bohr}$ with the Bohr radius $a_{Bohr} = \hbar^2/(me^2)$ it becomes: $\delta E_F = MgH$ equivalent to a potential which is $M/m$ times larger than our initial naïve estimate:

$$V' = MgH/e = (M/m)V, \qquad (14)$$

with a sign opposite to that of Eq. (10), clearly the $M/m$ enhancement appears also in the current generated. The electric field $E$ corresponding to Eq. (14) generates the following currents in our antenna [11]:

$$j = j_n + j_s = (\sigma_1 - i\sigma_2)E, \qquad (15)$$

where

$$\sigma_1 = (n_n e^2 \tau)/[m(1+\omega^2\tau^2)], \qquad (16)$$

$$\sigma_2 = n_s e^2/(m\omega) + n_n e^2(\omega\tau)^2/[m\omega(1+\omega^2\tau^2)]. \qquad (17)$$

Then, using the equation for the resulting electromotive force [1] it can be shown:

$$I = |jS| = [e\omega h_0 LS/(8m)]\left|\alpha_A n_s^A M_A - \alpha_B n_s^B M_B\right|, \qquad (18)$$

where $n_s^A$ and $n_s^B$ are the number densities of the superconducting electrons, and $\alpha_A$ and $\alpha_B$ are dimensionless constants typically of order ~0.1-1, which depend on the materials $A$ and $B$. In the above equations $\tau$ is the relaxation time of electrons due to collisions, typically ~$10^{-13}$ sec. These equations illustrate the great advantage of using superconductors: the current is larger by a factor of $1/\omega\tau \sim 10^{10}$.

    The above heuristic argument ignored the complexity of condensed matter systems, the band structure, the lattice symmetry and other aspects. All of these were discussed in detail in [12,13]. In addition, the relevant work functions for various materials and for different deformations may be calculated using band structure theory. Ab-initio codes, like QuantumWise, can numerically implement this. Using the above-mentioned code we calculated $\delta E_F/\delta l$ in Eq.(13) for various materials. This will be discussed in more detail elsewhere.

### VII. The improved signal, some comments and conclusions

    The large $M/m$ factor and the removal of the fudge factor $f$ at the end of the last paragraph jointly enhance the current induced by the GW relative to out naïve estimate by a factor of up to ~$10^6$. Thus even a $10^{-3}$ reduction due to residual magnetic blocking effects will leave us with a thousand fold enhanced signal – namely, $4 \cdot 10^{-13}$ Ampere which can be readily detected.

Constructing an antenna of the type envisioned here presents many technological challenges such as the proper implementation of the sub-micron spaghetti like conductors and their appropriate braiding. Also new unexpected difficulties and sources of noise may be discovered as we try to implement the ambitious program we have outlined here.

At the present time the LIGO project into which a monumental amount of resources and human efforts and ingenuity have been channeled and which has been running and debugged for some years by now is clearly most likely to discover GW. We believe that other approaches to the detection of GW – such as the use of the new technology of cold atoms [14,15] and also the highly developed SC technology that we are suggesting are worth pursuing.

Our device is also an antenna for ordinary electromagnetic waves and extreme care should be taken to minimize the effect of these. Thus an appropriate design with square symmetry kills the response to dipole radiation leaving only the far weaker higher multipoles. The same quadrupolar design also cancels the response to the dipole gravity field $g$ even if $g=g(t)$.[3]

The moderate size of the antenna allows many further refinements which are hard to implement in several *km* long designs as in LIGO. Completely surrounding it with a metallic film or multi-layer net forming an ideal Faraday cage will isolate it from any electromagnetic waves in the sub kHz region. Further isolation from high-energy cosmic ray particles can be achieved by putting it underground. Cooling to sub-Kelvin temperatures or putting it on a satellite reduces thermal or seismic noise respectively. If a periodic source is found, the antenna can be made to track it.

The single largest advantage is that by having several copies of the device at different terrestrial and space locations we can at any given time shield better some of the possible sources of noise. Even an extremely weak coincident signal in all of them will strongly suggest a GW source. This will be all the more the case if this happens in coincidence with a signal in LIGO which has completely different systematics and also with the other expected gamma ray bursts or neutrinos signals.

*Acknowledgements*. We would like to thank P. Shawhan for many insightful discussions of LIGO, and S. Davis for his comments. We are grateful to E. Flanagan and W. Walter for encouragement.

**References**

[1] A. Gulian, J. Foreman, V. Nikoghosyan, S. Nussinov, L. Sica and J. Tollaksen, "Spaghetti" design for gravitational wave superconducting antenna, *J. Phys.: Conf. Ser.* **507** (2014) 042013-4.
[2] A. Gulian, J. Foreman, V. Nikoghosyan, S. Nussinov, L. Sica and J. Tollaksen, Superconducting Antenna Concept for Gravitational Wave Radiation, http://arxiv.org/abs/1111.2655
[3] P.A.R. Ade *et al.* (BICEP2 Collaboration) Detection of B-mode polarization at degree angular scales by BICEP2, Phys. Rev. Lett. 112 (241101-23) 2014.
[4] http://www.nobelprize.org/nobel_prizes/physics/laureates/1993/
[5] B.P. Abbott *et al.* Searches for gravitational waves from known pulsars with Science Run 5 LIGO data, ApJ., **713** (2010) 671.
[6] M. Coughlin and J. Harms. Constraining the gravitational wave energy density of the Universe using Earth's ring. arXiv:1406.1147v1 [gr-qc] 4 June 2014.

---

[3] We would like to thank S. Davis for emphasizing this point.

[7] R.J. Adler, Long conductors as antennae for gravitational radiation. Nature **259** (1976) 296-297.
[8] G.M. Harry, Advanced LIGO: the next generation of gravitational wave detectors. Class. and Quantum Grav. 27, 084006.
[9] D. G. Blair, E.J. Howell, L. Ju, C. Zhao (Eds.), Advanced gravitational wave detectors. Cambridge University Press, Cambridge, 2012, p. 317.
[10] S.Weinberg, Gravitation and Cosmology, John Wiley and Sons, New York, 1972, Chapter 10.
[11] T. Van Duzer, C.W. Turner, Principles of superconducting devices and circuits. Prentice Hall, New Jersey, 1998, pp. 5.
[12] A.J. Dessler, F.C. Michel, H.E. Rorschach, G.T. Trammel, Gravitationally induced electric fields in conductors. Phys. Rev. 168 (1968)737-743.
[13] Sh.M. Kogan, Does an electron fall in a metallic pipe? Usp. Fiz. Nauk 105, 157-161,1971 [1972. Sov. Phys. Uspekhi 14, 658-661].
[14] R.D. Andersen, Suppose you want to glimpse the beginning of time, the very first moments of cosmic creation. Scientific American, 309, 42-47, 2013.
[15] A. Arvanitaki and A. A. Geraci. Detecting high frequency gravitational waves with optically levitated sensors. Phys. Rev. Lett 110, 071105, 2013.